\documentclass[aps,pra,twocolumn,showpacs,preprintnumbers,amsmath,amssymb,footinbib]{revtex4}
\usepackage{mathtools}

\DeclareMathOperator{\sinc}{sinc}
\DeclareMathAlphabet{\mathcalligra}{T1}{calligra}{m}{n}
\DeclareMathAlphabet{\mathpzc}{OT1}{pzc}{m}{it}

\usepackage{graphicx,epsfig}
\usepackage{bm}
\usepackage{dcolumn}
\usepackage{xcolor}
\usepackage[breaklinks=true,colorlinks,citecolor=blue,linkcolor=blue,urlcolor=blue]{hyperref}

\def\prv#1#2#3{Phys. Rev. {\bf #1}, #2 (#3)}

\def\ajp#1#2#3{Am. J. Phys. {\bf #1}, #2 (#3)}
\def\ejp#1#2#3{Eur. J. Phys. {\bf #1}, #2 (#3)}

\def\noi{\noindent}
\def\bc{\begin{center}}
\def\ec{\end{center}}
\newcommand{\bea}{\begin{equation}}
\newcommand{\eea}{\end{equation}\noi}
\newcommand{\ber}{\begin{eqnarray}}
\newcommand{\eer}{\end{eqnarray}\noi}
\begin{document}
\title{Rayleigh-Sommerfeld scalar diffraction by apertures moving at relativistic speeds}

\author{Rhitabrata Bhattacharyya$^{1}$}
\author{Saugata Bhattacharyya$^{2}$}
\author{Shyamal Biswas$^{3}$}\email{sbsp [at] uohyd.ac.in}
\affiliation{$^{1}$Central GST \& Central Excise, Kolkata North Commissionerate, Kolkata-700107, India\\ 
$^{2}$Department of Physics, Vidyasagar College, 39 Sankar Ghosh Lane, Kolkata-700006, India\\
$^{3}$School of Physics, University of Hyderabad, C.R. Rao Road, Gachibowli, Hyderabad-500046, India}

\date{\today}

\begin{abstract}
We have analytically obtained the theoretical results for the Rayleigh-Sommerfeld (R-S) scalar diffraction by apertures, such as single-slit, double-slit, grating and circular aperture, moving at relativistic speeds with the velocities perpendicular to the direction of incidence. We also have studied diffraction by a single-slit of oscillatory shutter. Our study would be significant in probing the relativistic transverse Doppler effect on the intensity pattern of the diffracted field.
\end{abstract}


\maketitle    

\section{Introduction}
Diffraction of electromagnetic wave by an aperture (e.g. single-slit, double-slit, diffraction grating, circular aperture, \textit{etc}) or an obstacle (e.g. circular disc, sphere, \textit{etc}) still continues to be a subject of great interest as applications of Huygens-Fresnel (H-F) principle in both the basic science \cite{Sommerfeld2,Born,Lahiri} and the applied science \cite{Goodman}. Fresnel-Kirchhoff (F-K) theory of scalar diffraction, which works for both the radiative near field region (Fresnel diffraction) and the far field region (Fraunhofer diffraction), is often considered as the theory for the diffraction of the monochromatic electromagnetic wave by an aperture (of size a few or more times larger than that of the wavelength of the electromagnetic wave) in non-dispersive medium, and is often applied for most of the practical purposes\cite{Born,Buchwald,Goodman,Lahiri} including the case of x-ray scattering by a crystal \cite{Bragg}. However, while F-K theory of scalar diffraction is not manifestly consistent, Rayleigh-Sommerfeld (R-S) theory \cite{Sommerfeld2} of scalar diffraction is manifestly consistent except at the edge of the aperture/obstacle \cite{Mukunda,Marchand}. Hence, we adopt Rayleigh-Sommerfeld scalar diffraction theory for our analysis at the observation point away from the aperture.  

Rayleigh-Sommerfeld obtained two solutions to the Helmholtz equation for the scalar field. The solutions are known as R-S diffraction integrals of the 1st kind (RSI-1) and the 2nd kind (RSI-2) which are compatible with the Kirchhoff boundary conditions of the 1st kind and the 2nd kind respectively. RSI-1 fits better than both the RSI-2 and the FKI in the far field region for oblique incidence \cite{Zurak}. Hence, we take RSI-1 for our analysis. However, hardly any literature, except the theoretical work done in the Ref.\cite{Dossou}, is found for studying diffraction of light by a moving aperture. Such a study is important for analysing Doppler effect on the intensity pattern for the diffraction. The authors of the Ref.\cite{Dossou}, however, analysed the non-relativistic longitudinal Doppler effect on the intensity pattern by adopting F-K scalar diffraction theory in this regard. In this article, we want to analyse the relativistic transverse Doppler effect by adopting R-S scalar diffraction theory \cite{Sommerfeld2}. 

For any relativistic description time appears as a coordinate.  Hence, we are bringing time ($t$) into the discussion unlike the case where the aperture is stationary and the description is time-independent. Naturally, the Green's function which appears in the RSI-1 would now become time-dependent. Let us consider a time-dependent monochromatic scalar field $\psi(\vec{r},t)=U(\vec{r})\text{e}^{-i\omega t}$ which satisfies the wave-equation
\begin{eqnarray}\label{eq0}
\Big(\frac{1}{c^2}\frac{\partial^2}{\partial t^2}-\nabla^2 \Big)\psi(\vec{r},t)=0
\end{eqnarray}
and results the Helmholtz equation
\begin{eqnarray}\label{eq1}
(\nabla^2+k^2)U(x,y,z)=0
\end{eqnarray}
for the time-independent part $U(\vec{r})=U(x,y,z)$ defined in a region $\mathcal{R}$ in the vacuum-space for the wavenumber $k=\omega/c$, (angular) frequency $\omega$ and the phase-velocity $c$ in the vacuum-space. If the monochromatic wave incidents from the left half-space $z<0$ and is diffracted in the right half-space $z>0$ by an aperture $\mathcal{A}$ of an opaque screen situated in the $xy$-plane, then the scalar field $U(\vec{r})$ at the point $\vec{r}=x\hat{i}+y\hat{j}+z\hat{k}$ in the right half-space $z>0$ is given by the RSI-1 \cite{Sommerfeld2,Born}
\begin{eqnarray}\label{eq3}
U(x,y,z)=\frac{1}{4\pi}\iint_{\mathcal{A}}U(x_0,y_0,0)\bigg[\frac{\partial G_-(\vec{r}) }{\partial z_0}\bigg]_{z_0=0}\text{d}x_0\text{d}y_0
\end{eqnarray}
where $R_+=+[(x-x_0)^2+(y-y_0)^2+(z-z_0)^2]^{1/2}$, $R_-=+[(x-x_0)^2+(y-y_0)^2+(z+z_0)^2]^{1/2}$ and $G_-(\vec{r})=\frac{\text{e}^{ikR_+}}{R_+}-\frac{\text{e}^{ikR_-}}{R_-}$ is an auxiliary function which is a combination of two time-independent Green's functions each of which obeys the Helmholtz equation along-with the scalar field $U(\vec{r})$ for $z_0\rightarrow0$. Thus, from Eqn.(\ref{eq3}) we get the time-dependent diffracted scalar field in the right half-space,  as
\begin{eqnarray}\label{eq4}
\psi(\vec{r},t)=\frac{\text{e}^{-i\omega t}}{4\pi}\iint_{\mathcal{A}}U(x_0,y_0,0)\bigg[\frac{\partial G_-(\vec{r})}{\partial z_0}\bigg]_{z_0=0}\text{d}x_0\text{d}y_0.
\end{eqnarray}
If the aperture moves uniformly with the velocity $v\hat{j}$ along the $y$-axis, then $y_0$ so as $R_+$, and $R_-$ change with time. This results the auxiliary function $G_-(\vec{r})$ in Eqn.(\ref{eq4}) to be time-dependent. Relativistic covariance further has to be employed in the wave equation (Eqn.(\ref{eq0})) for analysing the relativistic transverse Doppler effect on the diffracted scalar field.

Calculation in our article begins with the relativistic generalization of the wave equation (i.e. Eqn.(\ref{eq0})) and its solution (i.e. Eqn.(\ref{eq4})) in the both the rest fame and in the moving frame. Here by moving frame we mean -- the frame attached to the moving aperture, and by rest frame we mean -- the frame at rest with the source of the scalar field and the observation point for the scalar diffraction. Then we obtain the diffracted fields in the moving frame for (i) rectangular aperture, (ii) circular aperture, and (iii) diffraction grating. Then we obtain the diffracted field in the rest frame for the oscillatory shutter. We analyse a few of the results for both the radiative near field zone and the far field zone.  Finally, we discuss and conclude.

\section{Rayleigh-Sommerfeld scalar diffraction by moving apertures}
We apply H-F principle for the scalar diffraction in the right half-space ($z>0$) by the moving aperture ($\mathcal{A}$) which is lying in the $xy$-plane and moving uniformly with the velocity $v\hat{j}$ ($v<c$) along the $y$-axis as described before. The application goes in the line of the R-S scalar diffraction theory for the 1st kind. For simplicity, let us consider the scalar diffraction of a plane wave from the  moving aperture. The frame having the Cartesian coordinates $x$, $y$ and $z$ and the origin $O$, is  called as the rest frame, and is denoted as $S$. Now we need at least one moving frame for showing covariance of physical laws in the domain of the special theory of relativity. Let an aperture fixed frame having the Cartesian coordinates $x'$, $y'$ and $z'$ and the origin $O'$, be  called as the moving frame, and is denoted as $S'$. The primed and unprimed axes (i.e. $\vec{x}||\vec{x}'$, $\vec{y}||\vec{y}'$ and $\vec{z}||\vec{z}'$) are always parallel in our set-up. While the time of an event in the space-time is recorded as $t$ in the rest frame, the same is recorded as $t'$ in the moving frame. Let the respective clocks be initially synchronized ($t=t'=0$) for an event at $\vec{r}=\vec{r}'=0$. This implies that, both the origins of $S$ and $S'$ frames were coincident at $t=t'=0$. The primed coordinates follow from the unprimed coordinates through the Lorentz transformation. We further consider normal incidence of the plane wave $B\text{e}^{i(kz-\omega t)}$ with the 4-wave vector $(0,0,k,\omega/c)$ on the moving aperture in the rest frame.

\begin{figure}
\includegraphics[width=.98 \linewidth]{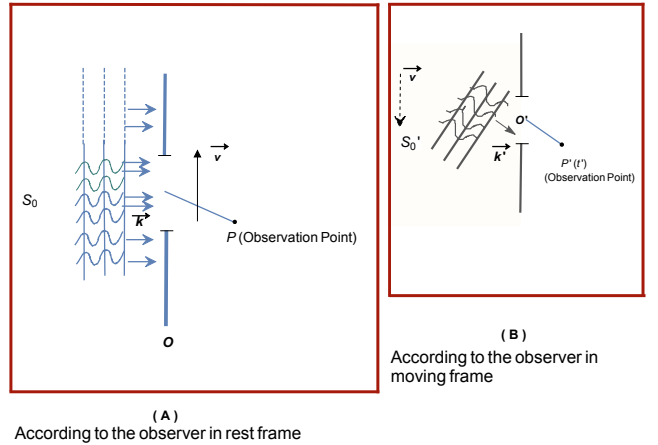}
\caption{Cartoons for respective observation of the wavefronts and subsequent diffraction. Here $O$ and $O'$ represent the origins of the rest frame and the moving frames respectively. Similarly, $\vec{k}$ and $\vec{k}'$ represent the wave-vectors in the rest frame and the moving frame respectively. While $S_0$ and $S_0'$ represent sources of the wavefronts in the rest frame and the moving frame respectively, $P$ and $P'$ represent observation point in the rest frame and the moving frame respectively.
\label{fig3}}
\end{figure}
  
\begin{figure}
\includegraphics[width=.98 \linewidth]{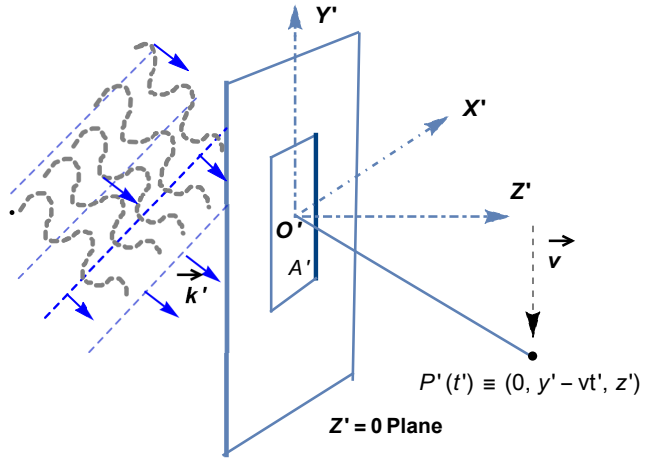}
\caption{Cartoon for the diffraction by the rectangular aperture in the moving frame. Here $A'$ is the area of the aperture in the moving frame.
\label{fig2}}
\end{figure}

Eqn.(\ref{eq0}) is already relativistically covariant. Thus, the wave equation is read in the moving frame, as
\begin{eqnarray}\label{eq5}
\Big(\frac{\partial^2}{c^2\partial t'^2}-\nabla'^2\Big)\psi'(\vec{r}',t')=0.
\end{eqnarray}
The moving observer, however, must take the time-dependent diffracted scalar field ($\psi(\vec{r}',t')$) to have the  same form as in Eqn.(\ref{eq4}) according to the principle of special relativity. Though the incident plane wave falls normally on moving aperture in the rest frame, it falls obliquely on the aperture in the moving frame with the 4-wave vector $(k_x',k_y',k_z',\omega'/c)$, where $k_x'=0$, $k_y'=[0-(v/c)(\omega/c)]\gamma$, $k_z'=k$, $\omega'=\gamma\omega$ and $\gamma$ is the Lotentz factor $1/\sqrt{1-v^2/c^2}$ \cite{Jackson}. This transformation leads to the transverse Doppler effect in the moving frame \cite{Jackson}.  Thus, while the  time-dependent incident scalar field at the aperture takes the form $\psi^{(i)}(x_0,y_0,0,t)=B\text{e}^{-i\omega t}$ in the rest frame, the same takes the form  $\psi'^{(i)}(x_0',y_0',0,t')=B\text{e}^{-i[\frac{v\omega'}{c^2}y_0'+\omega't']}$ in the moving frame. This clearly indicates -- while the phase at a given time is uniform at the aperture in the rest frame, it is uniformly increasing from its upper end to the lower end in the moving frame, as illustrated in FIG. \ref{fig3}.

However, if the field point is $(\vec{r},t)$ in $S$ frame and $(\vec{r}',t')$ in $S'$ frame, where the diffracted scalar field is observed, then, the causality demands   $c(t'-t_0')=R_+'(\vec{r}',t';\vec{r}_0', t_0')=R_-'(\vec{r}',t';\vec{r}_0', t_0')$ where $t_0'$ is earlier than $t'$ and $R_+'(\vec{r}',t';\vec{r}_0', t_0')=+[(x'-x_0')^2+(y'(t')-y_0(t_0'))^2+(z'-z_0')^2]^{1/2}$ is the distance between the object point ($\vec{r}'$) and secondary source point $(x_0',y_0',z_0')$ in the aperture for $z_0'\rightarrow0$, and $R_-'(\vec{r}',t';\vec{r}_0', t_0')=+[(x'-x_0')^2+(y'(t')-y_0(t_0'))^2+(z'+z_0')^2]^{1/2}$ is the distance between the object point ($\vec{r}'$) and the corresponding image point $(x_0',y_0',-z_0')$ at the aperture for $z_0'\rightarrow 0$. While the observation point moves in the moving frame, the secondary source point moves in the rest frame. Eventually, time of incidence of the scalar field on the aperture is not fixed for fixed time of observation in either of the frames. This allows interference of the diffracted fields of different time of incidence on the aperture in both the moving frame and the rest frame. Thus, we get the diffracted scalar field in the right half-space ($z'>0$) in the moving frame by generalizing the form of the Eqn.(\ref{eq4}), as
\begin{eqnarray}\label{eq6}
\psi'(\vec{r}',t')& &=\frac{1}{4\pi}\iint_{\mathcal{A}}\text{d}x_0'\text{d}y_0'\int_{-\infty}^{\infty}\text{d}t_0' \Bigg[\psi^{'(i)}(\vec{r}_0',t_0')\times \nonumber\\&&\bigg[\frac{\partial G_-'(\vec{r}',t'; \vec{r}_0',t_0')}{\partial z_0'}\bigg]_{z_0'=0}\Bigg]
\end{eqnarray}
where $\psi'^{(i)}(\vec{r}_0',t_0')=B\text{e}^{-i[\frac{v\omega'}{c^2}y_0'+\omega't_0']}$ is the incident scalar field at the aperture as evaluated above, and
\begin{eqnarray}\label{eq7}
G_-'(\vec{r}',t'; \vec{r}_0',t_0')&=&\frac{\text{e}^{i[k'R_+'-\omega'(t'-t_0')]}\delta(t'-t_0'-\frac{R_+'}{c})}{R_+'(\vec{r}',t';\vec{r}_0', t_0')}\nonumber\\&&-\frac{\text{e}^{ik'[R_-'-\omega'(t'-t_0')]}\delta(t'-t_0'-\frac{R_-'}{c})}{R_-'(\vec{r}',t';\vec{r}_0', t_0')}~~~~
\end{eqnarray}
is the auxiliary function where the $\delta$-functions take care of the causality constraint once the integration over $t_0'$ is evaluated. Each of the two parts of the auxiliary function is a retarded Green's function which obeys the wave equation (Eqn.(\ref{eq5})) for $z_0'\rightarrow0$ and $t'>t_0'$. Eqn.(\ref{eq6}) can be further recast after evaluating the derivative at $z_0'=0$ and integrating over $t_0'$, as
\begin{eqnarray}\label{eq6b}
\psi'(\vec{r}',t')&=&-\frac{z}{2\pi}\iint_{\mathcal A}\, dx_0'\, dy_0'\psi^{(i)}(\vec{r}_0',t'-R'(t')/c)\nonumber\\
&&\times\frac{1}{R'^2(t')}\bigg[ik'-\frac{1}{R(t')}\bigg]\nonumber\\&=&-\frac{zB\text{e}^{-i\omega't'}}{2\pi}\iint_{\mathcal{A}}\text{e}^{i[-\frac{vk'y_0'}{c}+k'\sqrt{x_0'^2+(vt'+y_0'-y')^2+z^2}]}\nonumber\\&&\times\bigg[\frac{ik'}{(x_0'^2+(vt'+y_0'-y')^2+z^2)}\nonumber\\&&-\frac{1}{(x_0'^2+(vt'+y_0'-y')^2+z^2)^{3/2}}\bigg]\text{d}x_0'\text{d}y_0'
\end{eqnarray}
where $R'(t')=R_+'|_{z_0'=0}=R_-'|_{z_0'=0}=+[(0-x_0')^2+([y'-vt']-y_0')^2+(z-0)^2]^{1/2}$, $\omega'=ck'$, $y'- vt'=y'(t')$ and $y_0'=y_0'(t')$. 

On the other hand, the time-dependent diffracted scalar field in the rest frame would be of the same form as in Eqn.(\ref{eq6}), as
\begin{eqnarray}\label{eq6a}
\psi(\vec{r},t)&=&\frac{1}{4\pi}\iint_{\mathcal{A}}\int_{-\infty}^{\infty}\psi^{(i)}(\vec{r}_0,t_0)\bigg[\frac{\partial G_-(\vec{r},t; \vec{r}_0,t_0)}{\partial z_0}\bigg]_{z_0=0}\nonumber\\&&\text{d}t_0\text{d}x_0\text{d}y_0\nonumber\\&=&-\frac{z}{2\pi}\iint_{\mathcal A}\text{d}x_0 \text{d}y_0\int_{-\infty}^{\infty}\, \text{d}t_0\psi^{(i)}(\vec{r}_0,t_0)\times\nonumber\\
&&\frac{\text{e}^{i[kR(t_0)-\omega (t-t_0)]}}{R^2(t_0)}\bigg[ik-\frac{1}{R(t_0)}\bigg]\delta(t-t_0-\frac{R(t_0)}{c})\nonumber\\
\end{eqnarray}
where $\psi^{(i)}(\vec{r}_0,t_0)=B\text{e}^{-i\omega t_0}$ is the normally incident scalar field at the moving aperture and the auxiliary function $G_-(\vec{r},t; \vec{r}_0,t_0)$ follows the same definition as that in Eqn.(\ref{eq7}) for the unprimed variables including $k$ and $\omega$, and $R(t_0)=R_+\big|_{z_0=0}=R_-\big|_{z_0=0}=+[(0-x_0)^2+(y-[vt_0+y_0])^2+(z-0)^2]^{1/2}$. Eqn.(\ref{eq6a}) is our key equation for the time-dependent diffracted field in the right half space ($z>0$) within the R-S scalar diffraction theory of the 1st kind by an aperture moving with a uniform relativistic speed perpendicular to the direction of incidence. Eqns.(\ref{eq6}) and (\ref{eq6b}) are also the key equations and serve the same purpose in the moving frame.  Using the covariance of the solution and the extra advantage offered by the moving frame in terms of  time independent locations of secondary source points, we obtain the intensity pattern in the moving frame for different kind of apertures. 

\subsection{Diffracted scalar field in the moving frame for a single slit}
\subsubsection{Rectangular aperture}
\label{Sec2A1}
The scalar diffraction in the right half-space ($z'=z>0$) by the rectangular aperture in the moving frame is described in the FIG. \ref{fig2}. The $O'X'Y'Z'$ frame is attached at the center of the aperture $\mathcal{A}$ such that $\{x_0',y_0',0\}\in\mathcal{A}$ for $-a/2<x_0'<a/2$ and $-b'/2<y_0'<b'/2$. While the wavefront of the plane wave of wavelength $\lambda'=\frac{2\pi}{k'}$ of the scalar field is obliquely incident on the aperture, the point of observation ($0,-vt'+y',z$), which though is fixed at ($0,y,z$) in the rest frame, is uniformly moving in the moving frame. Now, we evaluate Eqn.(\ref{eq6b}) for the rectangular aperture. The integral, however, takes a simple form for both the Fresnel zone ($\frac{[\text{inf}(a,b')]^4}{8z^3\lambda'}\ll1$) and the Fraunhofer zone ($\frac{[\text{inf}(a,b')]^2}{2z\lambda'}\ll1$) in the short time scale ($[vt'/z]^2\ll1$) and length scale ($(b'/z)^2\lesssim(y'/z)^2\ll1$) of observation. We employ above approximation on the distance between a secondary source point in the aperture and the moving observation point, and get
\begin{eqnarray}\label{eq7b} 
R'(t')&=&+[x_0'^2+([vt'-y']+y_0')^2+z^2]^{1/2}\nonumber\\
&=&\bar{z}(t')\bigg[1+\frac{2y_0'[vt'-y']-2y'vt'}{\bar{z}^2(t')}+\frac{x_0'^2+y_0'^2}{\bar{z}^2(t')}\bigg]^{1/2}\nonumber\\
&\simeq&\bar{z}(t')\bigg[1+\frac{y_0'[vt'-y']-y'vt'}{\bar{z}^2(t')}+\frac{x_0'^2+y_0'^2}{2\bar{z}^2(t')}\bigg]
\end{eqnarray}
where $\bar{z}(t')=+\sqrt{z^2+(vt')^2+y'^2}$ can be further approximated to be as $\bar{z}(t')\simeq z$ to obtain
\begin{eqnarray}\label{eq8}
\psi'(\vec{r}',t')&\simeq&\frac{-Bz\text{e}^{i[k'\bar{z}(t')-\omega't']}}{2\sqrt{2\pi}k'\bar{z}(t)}\text{e}^{-i[\frac{\pi}{4}+\frac{k'\bar{z}(t')[-v+\frac{c(vt'-y')}{\bar{z}(t')}]^2}{2c^2}]}\nonumber\\&&\times\text{e}^{-i\frac{k'vt'y'}{\bar{z}(t')}}\bigg[\frac{a\text{e}^{\frac{ia^2k'}{8\bar{z}(t')}}}{\sqrt{\bar{z}^3(t')/k'}}-2\sqrt{\pi}ik'\nonumber\\&&\times\big[\text{C}\bigg(\frac{a\sqrt{k'}}{2\sqrt{\pi\bar{z}(t')}}\bigg)+i\text{S}\bigg(\frac{a\sqrt{k'}}{2\sqrt{\pi \bar{z}(t')}}\bigg)\big]\bigg]\nonumber\\&&\times\bigg[\text{erfi}\bigg(\frac{\text{e}^{\frac{i\pi}{4}}\big[2v[-1+\frac{ ct'}{\bar{z}(t')}]+\frac{(b'-2y')c}{\bar{z}(t')}\big]}{2\sqrt{2}c\sqrt{1/k'\bar{z}(t')}}\bigg)\nonumber\\&&-\text{erfi}\bigg(\frac{\text{e}^{\frac{i\pi}{4}}\big[2v[-1+\frac{ct'}{\bar{z}(t')}]-\frac{(b'+2 y')c}{\bar{z}(t')}\big]}{2\sqrt{2}c\sqrt{1/k'\bar{z}(t')}}\bigg)\bigg]
\end{eqnarray}
where $\text{C}$, $\text{S}$ and $\text{erfi}$ represent Fresnel cosine integral, Fresnel sine integral and imaginary error function respectively. Above integral is valid for short time scale ($[vt'/z]^2\ll1$) in both the radiative near field zone and far field zone, and in the far field zone ($\frac{[\text{inf}(a,b')]^2}{2z\lambda'}\ll1$) it takes the form  $\psi'(\vec{r}',t')\simeq-i\frac{ab'z}{\lambda'\bar{z}^2(t')}B\text{e}^{i[k'\bar{z}(t')-\omega't']}\big[\sinc\big(\frac{\pi b'[-v+ c(vt'-y')/\bar{z}(t')]}{\lambda'c}\big)\big]$ exhibiting the Fraunhofer diffraction for $b\ll a$. The scalar field $\psi'(\vec{r}',t')$ is often equated with one of the components of the electric field ($\vec{E}'(\vec{r}',t')$) or magnetic field ($\vec{B}'(\vec{r}',t')$) of the electromagnetic field \cite{Goodman}. The intensity pattern  for the diffracted scalar field, however, is dependent of the choice of the inertial frames as the absolute value of diffracted scalar field is not a Lorentz invariant rather $\frac{1}{2}\epsilon_0E'^2(\vec{r}',t')-\frac{1}{2\mu_0}B'^2(\vec{r}',t')$ is a Lorentz invariant\footnote{See page number 599 of Ref.\cite{Jackson} for the Lorentz invariance.}. However, the Fraunhofer diffraction formula for the single-slit with $y'/z(t')\simeq\sin(\theta')$, $(vt'/z)^2\ll1$ and $(y'/z)^2\ll1$ would now be read as $b'\big[\sin(\theta')-\frac{vt'}{\bar{z}(t')}+\frac{v}{c}\big]=n\lambda'$ where $n=0, \pm1, \pm2,...$. The relativistic Doppler effect, which is absent in non relativistic limit, appears as $v/c$ term in the formula and results in shifting the central peak of the diffraction pattern. The Fraunhofer diffraction formula as obtained in the moving frame also defines excess optical path difference over $b\sin(\theta)$ for the motion of the observation point as $\delta[b'\sin(\theta')]\simeq b'v[-\frac{1}{c}+\frac{t'}{\bar{z}(t)}]$ resulting a time-delay $\tau=\frac{[\delta b'\sin(\theta')]}{c}\simeq b'v[-\frac{1}{c^2}+\frac{t'}{c\bar{z}(t)}]$. Result (Eqn.(\ref{eq8})) for the single-slit, however, can be directly used to determine the diffracted scalar field in the moving frame for the diffraction grating.

\subsubsection{Circular aperture}
If we replace the rectangular aperture by a circular aperture of the radius $a'$  in the moving frame, then the distance between a secondary source point in the aperture and the moving observation point would follow the Eqn.(\ref{eq7b}) subject to the constraint ($0\le x_0'^2+y_0'^2\le a'^2$). Now we evaluate  Eqn.(\ref{eq6b}) for the circular aperture in circular polar coordinates ($r_0,\phi_0; x_0'=r_0\cos(\phi_0), y_0'=r_0\sin(\phi_0)$) and for the short-time ($(vt'/\bar{z}(t))$) observation at point $\vec{r}'\equiv(0,y',z')$ in the Fraunhofer zone ($\frac{a'^2}{2z\lambda'}\ll1$) as 
\begin{eqnarray}\label{eq8b}
\psi'(\vec{r}',t')\simeq-i\frac{zk'a'^2B\text{e}^{i[k'\bar{z}(t')-\omega't']}}{\bar{z}^2(t)}\frac{\text{J}_1\big(a'k'\big[-\frac{v}{c}+\frac{vt'-y'}{\bar{z}(t')}\big]\big)}{a'k'\big[-\frac{v}{c}+\frac{vt'-y'}{\bar{z}(t')}\big]}\nonumber\\
\end{eqnarray}
where $J_1$ represents Bessel function of the first kind of order $1$. This result can be easily generalized for the arbitrary observation point $\vec{r}'\equiv(x',y',z')$ with the consideration of no transverse Doppler effect along the $x'$-axis, as 
\begin{eqnarray}\label{eq8bad}
\psi'(\vec{r}',t')&\simeq&-i\frac{zk'a'^2}{\bar{z}^2(t)}B\text{e}^{i[k'\bar{z}(t')-\omega't']}\nonumber\\&&\times\frac{\text{J}_1\big(a'k'\sqrt{\big[-\frac{v}{c}+\frac{vt'-y'}{\bar{z}(t')}\big]^2+\big[\frac{x'}{\bar{z}(t')}\big]^2\big)}}{a'k'\sqrt{\big[-\frac{v}{c}+\frac{vt'-y'}{\bar{z}(t')}\big]^2+\big[\frac{x'}{\bar{z}(t')}\big]^2}}~~~~
\end{eqnarray}
where $\bar{z}(t')=+\sqrt{z^2+(vt')^2+y'^2+x'^2}$ can be further approximated to be as $\bar{z}(t')\simeq z$ in the Fraunhofer zone. This form exhibits Fraunhofer diffraction for $v\rightarrow 0$ as expected from the previous sub-subsection. Fraunhofer diffraction formula for the moving observation point can now be obtained for the location of the Bessel-zeros.

\begin{figure}
\includegraphics[width=0.98
\linewidth]{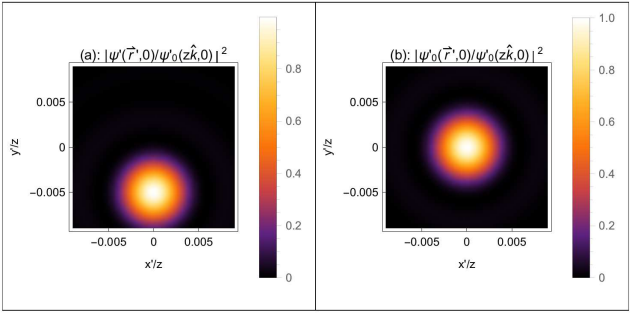}
\caption{(a): Density plot for the `intensity' $|\psi'(\vec{r}',0)|^2$ distribution on the screen at $z'=z$ in the $x'-y'$ plane for the scalar field diffracted by the circular aperture in units of the same ($|\psi_0'(z\hat{k},0)|^2$) obtained at $x'=0$, $y'=0$, $z'=z$, $t'=0$ for $v=0$. The diffracted field follows Eqn.(\ref{eq8bad}) for the parameters $t'=0$, $v/c=0.005$, $\lambda'/z=0.0001$ and $a'/z=0.01$.\\
(b): Density plot for the same `intensity' distribution for the same parameters except $v/c=0$.
}
\label{fig5a}
\end{figure}

We plot intensity distributions corresponding to Eqn.(\ref{eq8bad}) in the FIG. \ref{fig5a} for (a) $v/c=0.005$ and (b) $v/c=0$. The intensity distribution would resemble as the CCD camera image of the fringe patterns formed on an opaque screen at $z'=z$ in the $x'-y'$ plane. It is clear from the figure that, the transverse Doppler effect causes the fringes to shift opposite to the motion of the observer attached to the aperture. The amount of the shift ($\triangle y'$) follows from the location of the removable singularity in the $y'$ part of the $\psi'(\vec{r}',t')$ function in Eqn.(\ref{eq8bad}), and it results $\triangle y'=-\frac{v}{c}\bar{z}+vt'\simeq-0.005z$ for the parameters chosen for the FIG. \ref{fig5a} (a).

\subsection{Diffracted scalar field in the moving frame for the diffraction grating}
We now consider a diffraction grating, which is an array of single slits in the $x'y'$-plane, with the single slit spacing $d'$ along the $y'$ axis. We further consider that, there are $(N-1)/2$ single-slits along the +ve $y'$ axis and $(N-1)/2$ single-slits along the -ve $y'$-axis around the original one as considered in the previous subsection. The coordinate system remains unaltered with the original single silt in the diffraction grating. Middle of the $j$th slit with respect to that of the original one is separated by $y_j'=jd'$ where $j=0, \pm1, \pm2,..., \pm(N-1)/2$. Optical distance between a secondary source point in the $j$th aperture and the moving observation point $(0,-vt',z)$ is given by generalizing the form in Eqn.(\ref{eq7b}), as
\begin{eqnarray}\label{eq9} 
R'_j(t')&=&+[x_0'^2+([vt'-y']+[y_0'+jd'])^2+z^2]^{1/2}\nonumber\\
&\simeq&\bar{z}(t')\bigg[1+\frac{(y_0'+jd')[vt'-y']-y'vt'}{\bar{z}^2(t')}\nonumber\\&&+\frac{x_0'^2+(y_0'+jd')^2}{2\bar{z}^2(t')}\bigg]
\end{eqnarray}
where $\bar{z}(t')=+\sqrt{z^2+(vt')^2+y'^2}$ is further approximated to  $\bar{z}_j(t')\simeq z$ for a short time observation ($(vt'/z)^2\ll1$) for a small-sized ($j^2d'^2/z^2\ll1$ $\forall j$) grating. The approximation on $R_j'(t')$ would be useful for the scalar diffraction due to the grating in the short time scale in the far field zone. The diffracted scalar field at the observation point $\vec{r}'\equiv(0,y',z')$ is obtained as a superposition of contributions from individual single slits, which differ in their contribution by respective phases, is
\begin{eqnarray}\label{eq10}
\psi'(\vec{r}',t')&\simeq&\sum_{j=-(N-1)/2}^{(N-1)/2}-\frac{Bz}{\bar{z}(t)}\frac{\text{e}^{i[k'\bar{z}(t')-\omega't']}}{2\sqrt{2\pi}k'}\nonumber\\&&\times\text{e}^{-i[\frac{\pi}{4}+\frac{k'\bar{z}(t')[-v+c(vt'-y')/\bar{z}(t')]^2}{2c^2}]}\text{e}^{-ik'[\frac{vt'y'}{\bar{z}(t')}-\frac{jd'v}{c}]}\nonumber\\&&\times\bigg[\frac{a\text{e}^{\frac{ia^2k'}{8\bar{z}(t')}}}{\sqrt{\bar{z}^3(t')/k'}}-2\sqrt{\pi}ik'\big[\text{C}\bigg(\frac{a\sqrt{k'}}{2\sqrt{\pi\bar{z}(t')}}\bigg)\nonumber\\&&+i\text{S}\bigg(\frac{a\sqrt{k'}}{2\sqrt{\pi \bar{z}(t')}}\bigg)\big]\bigg]\nonumber\\&&\times\bigg[\text{erfi}\bigg(\frac{\text{e}^{\frac{i\pi}{4}}\big[2v[-1+\frac{ ct'}{\bar{z}(t')}]+\frac{(b'-2[ y'-jd'])c}{\bar{z}(t')}\big]}{2\sqrt{2}c\sqrt{1/k'\bar{z}(t')}}\bigg)\nonumber\\&&-\text{erfi}\bigg(\frac{\text{e}^{\frac{i\pi}{4}}\big[2v[-1+\frac{ct'}{\bar{z}(t')}]-\frac{(b'+2[y'-jd'])c}{\bar{z}(t')}\big]}{2\sqrt{2}c\sqrt{1/k'\bar{z}(t')}}\bigg)\bigg].~~~~~
\end{eqnarray}
Eqn.(\ref{eq10}) takes a simpler form in the far field zone ($\frac{[\text{inf}(a,b')]^2}{2z\lambda'}\ll1$) for $b'\ll a$ and $d'j\lesssim y'$ $\forall j$ in the short time scale ($(vt'/z)^2\ll1$), as
\begin{eqnarray}\label{eq11}
\psi'(\vec{r}',t')&\simeq&-i\frac{ab'z}{\lambda'\bar{z}^2(t')}B\text{e}^{i[k'\bar{z}(t')-\omega't']}\frac{\sin(\frac{\pi d'N[vt'-y']}{\bar{z}(t')\lambda'})}{\sin(\frac{\pi d'[vt'-y']}{\bar{z}(t')\lambda'})}\nonumber\\&&\times\bigg[\sinc\bigg(\frac{\pi b'[-v/c+[vt'-y']/\bar{z}(t')]}{\lambda'}\bigg)\bigg].~~~~~
\end{eqnarray}
The appearance of the $v/c$ term, accounts for the shift in wave number, as required by the relativistic transverse Doppler effect. One can easily get back the usual Fraunhofer diffraction result for the grating, as well as that for the single-slit ($N=1$) by setting $v\rightarrow0$ now. 

\begin{figure}
\includegraphics[width=.98 \linewidth]{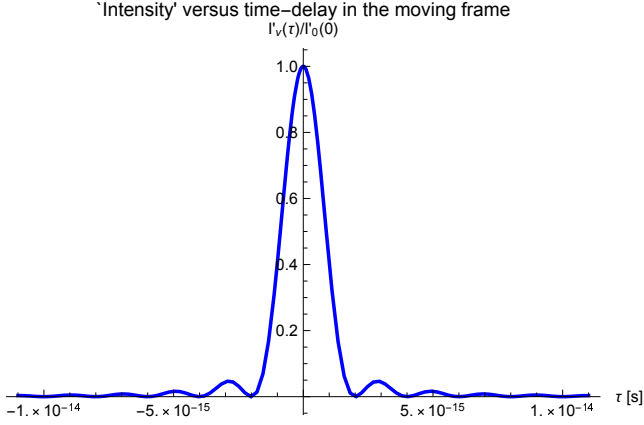}
\caption{Time-delay dependence of the `intensity' $I'_v(\tau)=|\psi'(0,0,z',0)|^2$ of the diffracted field in units of $I'_{0}(0)$ which represents Fraunhofer diffraction result for the `intensity' at any time at the point ($0,0,z'$). The diffracted field  follows Eqn.(\ref{eq11b}) for the parameters $c=3\times10^8~$m/s and $\lambda'=6000~$\AA.
\label{FIG-1}}
\end{figure}

We need to equate $\hbar k'$ with the magnitude of the momentum of an electron (keeping $v/c$ as it is for the relativistic effect) in the moving frame to apply Eqn.(\ref{eq11}) for the electron-diffraction. Diffraction due to the grating would look like that due to the single-slit (except scaling by $N$) for $t'\rightarrow0$ and $y'\rightarrow0$. The time-delay $\tau=b'v[-\frac{1}{c^2}+\frac{t'}{c\bar{z}(t')}]$ which was defined for the diffraction by the rectangular aperture would still be useful for the grating at least for $t'\rightarrow0$ and $y'\rightarrow0$. Eqn.(\ref{eq11}) in terms of the time-delay takes the form
\begin{eqnarray}\label{eq11b}
\psi'(0,0,z',0)&\simeq&-i\frac{ab'N}{\lambda'z}B\text{e}^{ik'z}\big[\sinc\big(\pi c\tau/\lambda'\big)\big]
\end{eqnarray}
for $t'\rightarrow0$ and $y'\rightarrow0$.  We plot modulus squared of the right hand side of Eqn.(\ref{eq11b}) with respect to the time-delay ($\tau$) at $y'=0$ and $t'=0$ in the FIG. \ref{FIG-1}. It is clear from the figure that, the coherence so as the relative intensity at a point decreases with the increase of the absolute value of the time-delay.

\begin{figure}
\includegraphics[width=0.98
\linewidth]{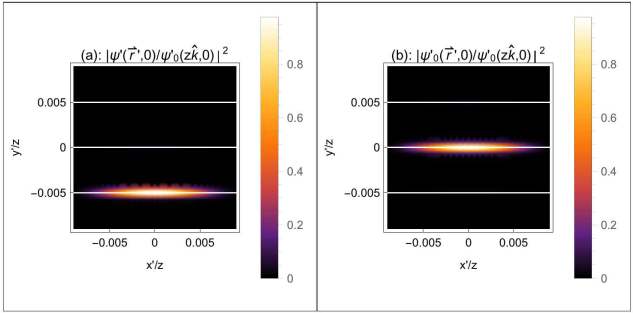}
\caption{(a): Density plot for the `intensity' $|\psi'(\vec{r}',0)|^2$ distribution on the screen at $z'=z$ in the $x'-y'$ plane for the scalar field diffracted by the grating in units of the same ($|\psi_0'(z\hat{k},0)|^2$) obtained at $x'=0$, $y'=0$, $z'=z$, $t'=0$ for $v=0$. The diffracted field follows Eqn.(\ref{eq11xy}) for the parameters $t'=0$, $v/c=0.005$, $\lambda'/z=0.0001$, $a/z=0.01$, $b'/z=0.05$, $d'/z=0.02$ and $N=5$.\\
(b): Density plot for the same `intensity' distribution for the same parameters except $v/c=0$.
}
\label{fig5b}
\end{figure}

Only the single-slit diffraction pattern would appear along the $x'$-axis if we consider $b'\lesssim a$ as because the transverse Doppler effect does not take place along the $x'$-axis by construction. Thus the scalar field at the observation point $\vec{r}'\equiv(x',y',z')$ in the far field zone can be obtained from Eqn.(\ref{eq11}) with the proper consideration of the diffraction along the $x'$-axis for the grating in the moving frame, as
\begin{eqnarray}\label{eq11xy}
\psi'(\vec{r}',t')&\simeq&-i\frac{ab'z}{\lambda'\bar{z}^2(t')}B\text{e}^{i[k'\bar{z}(t')-\omega't']}\frac{\sin(\frac{\pi d'N[vt'-y']}{\bar{z}(t')\lambda'})}{\sin(\frac{\pi d'[vt'-y']}{\bar{z}(t')\lambda'})}\nonumber\\&&\times\bigg[\sinc\bigg(\frac{\pi b'[v/c+[vt'-y']/\bar{z}(t')]}{\lambda'}\bigg)\bigg]\nonumber\\&&\times\bigg[\sinc\bigg(\frac{\pi b'x'/\bar{z}(t')]}{\lambda'}\bigg)\bigg]
\end{eqnarray}
where $\bar{z}(t')=+\sqrt{z^2+(v't)^2+y'^2+x'^2}$ can be further approximated to be as $\bar{z}(t')\simeq z$ in the far field zone. We plot intensity distributions corresponding to Eqn.(\ref{eq11xy}) in the FIG. \ref{fig5b} for (a) $v/c=0.005$ and (b) $v/c=0$ as we have done for the circular aperture. It is also clear from the figure that, the transverse Doppler effect causes the fringes to shift opposite to the motion of the observer attached to the aperture. Amount of the shift ($\triangle y'$) follow from the location of the removable singularity in the $y'$ part of the $\sinc$ function in Eqn.(\ref{eq11xy}). The function though is different from that for the circular aperture, the location of removable singularity interestingly remains same.

\subsection{Diffraction by single-slit of oscillatory shutter}
\begin{figure}
\includegraphics[width=0.98
\linewidth]{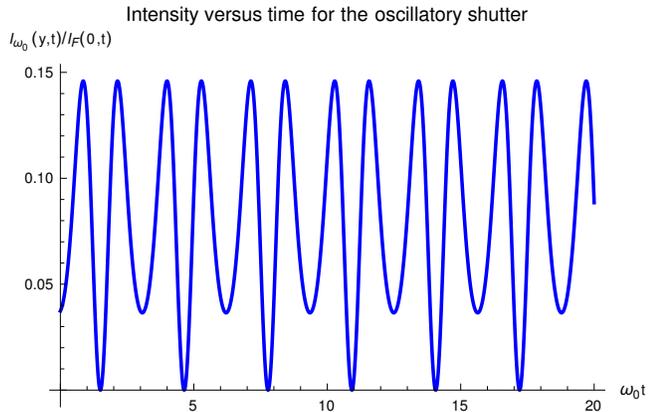}
\caption{Time-dependence of the `intensity' $I_{\omega_0}(y,t)=|\psi(0,y,z,t)|^2$ of the diffracted field in units of $I_{F}(0,t)$ which represents Fraunhofer single-slit diffraction result for the `intensity' at any time at the point ($0,0,z$).  The diffracted field  follows Eqn.(\ref{eq13}) for the parameters $c=3\times10^8$m/s, $\lambda=6000~$\AA, $z=0.5~$m, $y=z/50$, $b=0.00025~$m and $\omega_0=10^{15}~$Hz.
}
\label{fig4}
\end{figure}

We already have considered scalar diffraction by a rectangular aperture in the section \ref{Sec2A1}. Let us now study the same with the only difference that the slit is now being opened and closed in regular interval of time $2\pi/\omega_0$ along the $\hat{y}_0$ direction. The slit though as a whole is not moving, its edges are oscillating with same frequency in the rest frame such that $(-b/2)\sin \omega_0 t_0<y_0< (b/2)\sin \omega_0 t_0$ holds for this case. Diffracted scalar field at the observation point $\vec{r}\equiv(0,y,z)$ for this case, can be determined from Eqn.(\ref{eq6a}) for $R(t_0)=[(0-x_0)^2+(y-y_0)^2+(z-z_0)^2]^{1/2}$ $\forall t_0$. The scalar diffracted field takes simple form in the Fraunhofer zone ($\frac{[\text{inf}(a,b)]^2}{2z\lambda}\ll1$) once we approximate the following $R(t_0)\simeq\bar{z}[1-yy_0/\bar{z}^2]$ where $\bar{z}=+[z^2+y^2]^{1/2}$. Thus, we evaluate the integrations over $x_0$ and $y_0$, as
\begin{eqnarray}\label{eq12}
\psi(\vec{r},t)&\simeq&-i\int_{-\infty}^{\infty}\frac{abz}{\lambda\bar{z}^2}B\text{e}^{i[k\bar{z}-\omega t]}\big[\sinc\big(\frac{\pi yb\sin(\omega_0t_0)}{\lambda\bar{z}}\big)\big]\nonumber\\&&\times\sin(\omega_0t_0)\delta(t-t_0-\bar{z})\text{d}t_0,
\end{eqnarray}
where only in $\delta(t-t_0-R(t_0))$ we have approximated $R(t_0)\simeq\bar{z}$ for $\bar{z}^2\gg yb/2$. Now evaluating the time-integral in Eqn.(\ref{eq12}) we get diffracted scalar field at the point of observation ($0,y,z$), as
\begin{eqnarray}\label{eq13}
\psi(\vec{r},t)&\simeq&-i\frac{abz}{\lambda\bar{z}^2}B\text{e}^{i[k\bar{z}-\omega t]}\big[\sinc\big(\frac{\pi yb\sin(\omega_0[t-z/c])}{\lambda\bar{z}}\big)\big]\nonumber\\&&\times\sin(\omega_0[t-z/c]).
\end{eqnarray}
We plot modulus squared of the right hand side of Eqn.(\ref{eq13}) in the FIG. \ref{fig4}. The time $z/c$ taken by the scalar field to reach the point of observation from the slit, takes care of the relativistic effect on the scalar diffraction. The oscillation in the intensity pattern is coming from both the $\sinc$ function and the sinusoidal function.

\section{Conclusion}
To conclude, we have illustrated relativistic generalization of the Rayleigh-Sommerfeld scalar diffraction theory for various moving apertures such as rectangular aperture, circular aperture and diffraction grating. Our results for the moving apertures capture the transverse Doppler effect, i.e. increment of the angular frequency ($\omega'-\omega=\omega[\gamma-1]$) and that of the wavelength ($\lambda'-\lambda=\lambda[1/\gamma-1]$), once the intensity distribution of the diffracted fields are observed from the moving frames. Our results are fairly accurate for all possible values of speed of the moving apertures. We have considered only uniform motion of the apertures in the article. Consideration of non-uniform motion of the apertures would replace the special relativity by the general relativity.

While the experimental realization of the transverse Doppler effect on the scalar diffraction would be difficult, experimental realization of the scalar diffraction by the oscillatory shutter would be easy. The transverse Doppler effect, can however, be analogously observed, as time-delay of order $10^{-15}$s as shown in FIG. \ref{FIG-1} is easily measurable quantity now-a-days, by virtue of maintaining uniformly increasing phase shift from top to bottom of the aperture in the rest frame, e.g. by setting a prism next to the aperture and uniformly rotating it for some time interval.

Rayleigh-Sommerfeld scalar diffraction theory does not take aperture-edge effect into account \cite{Mukunda,Marchand}. Aperture-edge effect, however, is not significant if the dimension of the aperture is reasonably larger than the wavelength of the scalar field incident on it. Rayleigh-Sommerfeld scalar diffraction theory would be applicable for the diffraction of electromagnetic wave in this situation. Otherwise, we need to generalize the vector diffraction theory \cite{Vector-Diffraction} for the diffraction of electromagnetic wave by the moving apertures. We are keeping this as an open problem.

We have considered scalar field for mass-less particles. Scalar field, however, obeys the Klein-Gordon equation instead of the wave equation (Eqn.(\ref{eq5})) for scalar particles of nonzero mass. How to generalize the Rayleigh-Sommerfeld scalar diffraction theory from the Klein-Gordon equation is also kept as an open problem. 

Scalar diffraction theory is often applied to the diffraction of beam of electrons. This is possible if the wavelength associated with an electron in the beam is much smaller than the dimension of the aperture.  Otherwise, we need to generalize both the path-integral formalism of the quantum theory of diffraction for the Dirac spinors \cite{Feynman} and the quantum theory of diffraction in the light of quantum measurement \cite{Fabbro} for the uniform motion of the apertures. We are also keeping this as an open problem.

Finally, in our diffraction setup one can have either a moving slit or an oscillatory slit where one can see the interplay of length and time scales - a setting compatible with the ghost imaging experiments \cite{Moreau,Shih,Bornman}, Hong-Ou-Mandel interferometry \cite{Faccio} and the field of structured light \cite{Halina}. Our work will kindle interest in both the theoretical as well as experimental fronts.

\section*{Acknowledgement}
R. Bhattacharyya acknowledges hospitality of the School of Physics, University of Hyderabad, India while visited the institution in December, 2019. S. Biswas acknowledges partial financial support of the SERB, DST, Govt. of India under the EMEQ Scheme [No. EEQ/2019/000017]. Useful discussions with Dr. N. Sri Ram Gopal  (UoH, Hyderabad) and Prof. S. Dutta Gupta (UoH \& TCIS, Hyderabad) are gratefully acknowledged. We thank the anonymous reviewers for their thorough review, and highly appreciate their comments and suggestions which significantly contributed to improving the quality of the presentation.

\end{document}